\documentstyle[12pt]{article}

\begin{document}
\title{On the utility and implication of convex effective potential
 for the Higgs particles}
\author{Jifeng Yang \\
\it Department of Physics \\
\it and\\
\it School of Management \\
\it Fudan University, Shanghai, 200433, P. R. China}
%\date{\today}
\maketitle
\begin{abstract}
It is suggested that from an alternative point of view, the convex
full effective potential (EFP) might also be useful. A convex EFP
with spontaneous symmetry breaking (SSB) make the Higgs
modes (elementary or effective) complicated in that the normal
final states (experimentally identifiable) for Higgs modes
could not be analytically defined. This observation might have
some nontrivial consequences for the Standard
Model (SM) and particle physics.
\end{abstract}

PACS number(s): 11.30.Qc;12.10.Dm;14.80.Bn;14.80.Cp.
%\newpage
\vspace {1.5cm}

It is known that the Higgs sector effecting SSB of the
flavor symmetry has been a quite complicated part of the SM
inspite that SM has been deemed as well established by the
community of high energy physics (HEP)\cite {Alt}. To further
understand the SM and physics beyond SM, sufficient
investigations on the SSB and the Higgs sector or its equivalent
should be made. There can be found a number of attempts heading
for a deeper understanding of the complicated Higgs sector.
There is even modeling work trying to provide masses for
intermediate gauge bosons without introducing Higgs
model \cite {Wu}.

In this note, we attempt to suggest another aspect of the SSB
effected through scalar fields. We will adopt the effective
potential (EFP) approach \cite {Jack} which is well known as
a useful tool for studying SSB \cite {EFPSSB}.

Before starting our main arguments, we need to make some
remarks on the EFP that is also known as the generating 
functional for one-particle-irreducible(1PI) Green function(Gf)s 
\cite {Jack}. 

It is known that the full EFP is real and convex for any QFT
within which it can be consistently defined \cite {Syman}.
However, due to UV divergences, one might wonder if
renormalization could violate the convexity and there has
been a lot of literature investigating this impact
\cite {Renvex,Sher}. We will follow the standard point of
view that renormalization would not affect the convexity provided 
it is appropriately done \cite {Renvex}. In fact, if one would 
adopt the underlying theory approach recently proposed by the 
author to deal with the unphysical UV infinities \cite {Yang}, 
the convexity would naturally follow from that given by the 
postulated well defined underlying theory or equivalently from 
the consistent definitions of the radiatively arised constants
(the explicit demonstration of this point will be given in a 
more detailed report \cite {EPAconv}).

In practical calculations people
often arrive at nonconvex and/or complex effective potentials,
which seems to be in conflict with the above assertion. This 
issue has been discussed in literature \cite {Eric,Sher}.
The solution lies in that the full EFP is complex where it is
nonconvex. The imaginary part in fact arises if one starts
from perturbative framework of the EFP where the parameters
(masses, couplings,etc.) are first defined with nonconvex
lagrangian potential, or, the EFP is defined in such kind of
formulation that the field configurations (or states) there
defined are not all stable ones indicated by the imaginary part.
It is shown by Weinberg and Wu \cite {Eric}
that this imaginary part multiplied with the space volume is
half the decay constant of the unstable modes. Conventionally
it is held that the convex EFP is not the physically interested
one. In the case of the nonconvex lagrangian potential, it is
the orthodox point of view that the field parameters are not
expectation values of localized (homogeneous) states but that of
superposition of distinct states, i.e.,the convex EFP is not
given by homogeneous states everywhere \cite {Eric,Sher,Curt}

In this note, however, we follow a somewhat unorthodox line of
argument. It is known that the Landau model (nonconvex potential)
is an effective phenomenological model, not the genuine formulation
of the true physics. This formulation is quite economic--merely a
few coefficients of the polynomial potential in terms of the order
parameter are given. Then there are some ingredients in this
simplified model that are not thermodynamically stable. Similar
things happen in the van der Waals theory which is also an
effective formulation. Now we ask the following question: what
will the more complete (not just effective) and thermodynamically
stable formulations of the
phenomena effectively described by the Landau model and the
van der Waals model look like? First, it will certainly be more
complicated than the effective ones in that it can not be simply
given by the polynomial in terms of certain phenomenological
parameters. Second, no thermodynamically unstable modes or
ingredients should be inherent in the complete formulations. Third,
since the complete formulations is strictly thermodynamically stable,
the complete thermodynamical potentials must be convex \cite {RU},
in other words, the complete formulations should naturally yield
convex thermodynamical potentials in terms of thermodynamically
stable parameters. For quantum field theory case, referring to
Weinberg and Wu's discussion \cite {Eric}, {\sl it is
natural to ask what are the final outcome of the decay indicated by
the imaginary part we mentioned above?} For
the decay to stop and for the imaginary part to vanish, the
final outcome will be necessarily described by a real and
convex EFP defined in terms of stable states or field
configurations. In other words, if one could reformulate
the theory with stable modes and states, there should be
no imaginary part and nonconvex shape for the EFP at all,
and there is of course no obstacle to interpret this
EFP as minimum energy densities for homogeneous field
configurations in contrast to the conventional formulation with
unstable states inherent there. Thus, following the above unorthodox
argument, {\sl one does not need to resort to the superposition of two
'distinct' states to understand the flat bottom of the EFP
given in stable modes or field configurations.}

In thermodynamics Maxwell construction
procedure \cite {Huang} serves to remove the
thermodynamically unstable ingredients from the van der
Waals formulation, and the resulting formulation is only made
up of thermodynamically stable modes and describes
thermodynamics faithfully. The legendre transform
plays the same role in QFT--removing the unstable
ingredients dominated if any and leading to a formulation
with only quantum mechanically stable states or modes.
For a formulation without original unstable modes
(e.g., there is no nonconvex part in the lagrangian)
the legendre transform will not alter the contents at all
and just pick out the 1PI Gfs trivially. However, for the
same phenomenon both the original 'unstable' formulation
and the 'stable' one should, {\sl after legendre transform},
be physically the same and the convexity and reality of
the EFPs follows automatically.

There remains one issue to be addressed, the meaning of
the flat bottom. Since the EFP by definition should have
taken in all the quantum effects to define the generating
functions of the 1PI correlation functions for a QFT, this
region is in fact dynamically isolated from all the other
part of the field configuration space where the EFP is not
flat. (It is such a strange region that each point is
quantum mechanically isolated from the other points except
they are related by symmetry.) That means the flat region
is physically forbidden, or the nontrivial physical
phenomenon could in effect be meaningful only beyond this
region. This is a bit different from the orthodoxical
point of view that {\it the convex EFP is not infromative
about the quantum theory in this flat bottom region, but
the effective action is} \cite {Curt}. We wish to suggest
that we could
extract consistent information from the convex EFP in the case
with flat segement in EFP provided we treat it in an alternative
point of view. It is also in accord with the spirit of the use
of the vacuum manifold in string and SUSY literature.

Hence, {\sl to get a real and convex EFP for the full
quantum theory, the theory should be formulated in terms of
stable field configurations and states.} How to find such states
or equivalently the
appropriate parameters? It is a demanding job to find such
a formulation, especially for the Higgs physics that interests
us. For our purpose here,
we only need to assume here that we CAN formulate the full
quantum theory containing scalar sector effecting SSB and
feeding the gauge fields with masses in terms of well
defined effective parameters so that all the states and
modes are stable. The scalar sector will also be
named as the Higgs sector (which should be different
from the conventional one with nonconvex lagrangian
potentials \cite {Higg}). There should be no
nonconvex part in the lagrangian and the EFP
for the Higgs sector should be well defined everywhere
for homogeneous states. ( There might be a flat segment
in the lagrangian where the scalar fields could be viewed
as 'moduli fields', following the SUSY literature. It
would be an interesting try to start with a lagrangian
potential with flat bottom and construct its quantum theory).
                                            
We wish to point out that even with convex model, the
perturbative
truncations and or some nonperturbative approximation of
the EFP in practical calculations might be nonconvex
(however, without imaginary part necessarily presents
as long as the formulation excludes unstable states) while
the full EFP should always be real and convex. (see Ref.
\cite {Yang1} for an example). Of course the formulation
in which there are unstable states is not completely
useless but is DOMINATED by the stable formulation.
One can refer to the statistical mechanics \cite {Huang}
about the case of metastable thermodynamical states
(of course not most probable states) for an analogue.

All the above assertions are meaningful only when the
EFP approach can be consistently defined within a QFT
\cite {Curt}. We remind the reader that we will follow the
unorthodox point of view elaborated above in the subsequent
discussions. No arguments in this note are claimed to be
firmly established as the convexity is
conventionally disliked and less trusted. We only wish
to add to favor the use of the convex EFP in a modified
way. We feel that convex EFP is a bit more informative
about the physics than it is conventionally held.

It suffices for our purpose to focus on the Higgs sector
(we will not specify the space-time dimension as our
arguments here do not depend upon it). Then denoting
the full EFP by $U_{eff} (\phi)$ for the Higgs sector
(whatever kind, fundamental or composite, as long as
the EFP can be consistently defined for the sector and
the formulation is free of unstable states or modes ),
its convexity is expressed by the following inequality,
%%%%%%%%%%Eq(1)
\begin{equation}
\partial_{\phi_i} \partial_{\phi_j} U_{eff} ({\bf \phi})
   \geq 0, i,j=1, \cdots, N
\end{equation}
where ${\bf \phi}$($=(\phi_i)$) refers to the vacuum
expectation values of the scalar fields as a vector in
the N-flavor space in the Higgs model whose flavor
symmetry is spontaneously broken. (Note that from now
on all the parameters refer to those with which the
formulation of the theory is free of unstable states)
That is, $U_{eff}({\bf \phi})$ is invariant under the
action of the symmetry group $G_{flavor}$ while the
vacuum state $|0 \rangle$ is not,
%%%%%%%Eq(2)
\begin{equation}
\hat U (g) |0 \rangle \neq |0 \rangle\>, g \in G_{flavor}
\end{equation}
with $\hat U (g)$ denoting the unitary representation
of the group $G_{flavor}$ in QFT and 
${\bf \phi}_{vac}= \langle 0|\hat {\bf \phi}|0 \rangle (\neq 0)$ 
minimizes $U_{eff} ({\bf \phi})$:
%%%%%%%%Eq(3)
\begin{equation}
\partial_{\phi_i} U_{eff} ({\bf \phi}) =0, \ \
\partial_{\phi_i} \partial_{\phi_j} 
U_{eff} ({\bf \phi}) \geq 0
\end{equation}
in a small neighborhood of ${\bf \phi}_{vac}$, or equivalently
%%%%%%%Eq(4)
\begin{equation}
U_{eff} ({\bf \phi}) \geq U_{eff} ({\bf \phi}_{vac}), \ \ \ \
 \forall {\bf \phi}
\end{equation}
while the degeneracy of the vacuum states indicates the
existence of Goldstone modes \cite {Gold}.

Now combining Eqs. (3),(4) and (1), it is easy to see that
the EFP must have a flat bottom, i.e.,
%%%%%%Eq(5)
\begin{equation}
U_{eff} ({\bf \phi}) \equiv U_{eff} ({\bf \phi}_{vac}), \ \ 
\forall {\bf \phi} \in A:=\{{\bf \phi}: 
|{\bf \phi}|\leq |{\bf \phi}_{vac}|\}.
\end{equation}
Then obviously,
%%%%%Eq(6)
\begin{equation}
\Gamma_{i_1 \cdots i_n}:=
\partial_{\phi_{i_1}} \cdots \partial_{\phi_{i_n}} 
U_{eff} ({\bf \phi}) \equiv 0 ,\ \ \ \forall n \geq 1, 
{\bf \phi}\in A^0:= \{{\bf \phi}: |{\bf \phi}|
 < |{\bf \phi}_{vac}|\}
\end{equation}
while $\Gamma_{i_1 \cdots i_n}$ could not vanish identically
when ${\bf \phi}$ does not belong to the set $A^0$--including
the vacuum state. Since we know that at least at the
classical limit the EFP should have a positive two-point
1PI Gf--the would-be effective mass, then {\bf there must
be} an index $i_0$ such that
$\left. \partial_{\phi_{i_0}}^2 U_{eff}
\right |_{{\bf \phi}={\bf \phi_{vac}}}>0$. This fact in turn 
implies that the higher order($\geq 3$) derivatives of the
EFP with respect to ${\bf \phi}$ can {\bf NOT} be analytical
functions at the vacuum state. Thus we arrive at

{\bf Proposition I}\ \ \
{\sl The full effective potential for Higgs model effecting
SSB could
not be expanded into analytical Taylor series around the
vacuum state or any state degenerate with the vacuum.}

Since the EFP is at the same time the 1PI Gf generating
functional, the effective interactions can be found as
various partial derivatives of the EFP W.R.T. the
components of ${\bf \phi}$. Then it follows immediately
that excluding the gravitational interaction the sector
defined by the set $A^0$ is a totally isolated sector in that
each state (modulo degeneracy) in
this sector is totally isolated with any other one
(including states beyond $A^0$) due to the {\sl absence
of correlation functions in this sector}. The physical vacuum is
stable in the sense that it is isolated from this strange sector.
We {\bf do not need to employ a nonconvex and approximate EFP
to stablize the physical vacuum and hence the SM physics}.
This
isolated sector corresponds in fact to the
aforementioned region of scalar field parameter space
where the conventional perturbatively defined EFP
is nonconvex and complex. Here without the use of
superposition of localized states and nonconvexity we
also arrive at the same stability conclusion of the
physical vacuum. In fact we achieved more in the
{\sl stable} parameter formulation. Of course
SM could not stand on any state in this strange sector
but only be established on the physical vacuum state
${\bf \phi}_{vac}$ that could not transit into the
$A^0$ sector.

Back to our main track, from Proposition I and the
discussions above, we can not obtain the following
Taylor expansion:
%%%%%%Eq(7)
\begin{equation}
U_{eff} ({\bf \phi}_{vac}+\delta {\bf \phi}) =
U_{eff} ({\bf \phi}_{vac})+\frac {1}{2} 
(M_H^2(\phi_{vac}))_{ij} (\delta \phi)_i (\delta \phi)_j
+ R (\delta {\bf \phi})
\end{equation}
with $|\delta {\bf \phi}|/|{\bf \phi}_{vac}|$ being
sufficiently small, $(M_H^2(\phi_{vac}))_{ij}:=\left.
\partial_{\phi_i} \partial_{\phi_j} U_{eff}
\right |_{{\bf \phi}={\bf \phi}_{vac}}$ and $R(\cdots)$
denotes the residual terms that are even smaller
comparing to the second mass term. The reason is that
the effective mass matrix for Higgs modes is
discontinuous at ${\bf \phi}={\bf \phi}_{vac}$,
%%%Eq(8)
\begin{equation}
(M_H^2(\phi_{vac}^-))_{ij} (\equiv 0) \neq
(M_H^2(\phi_{vac}^+))_{ij} (>0)
\end{equation}
and hence the residual term in the Taylor expansion
is in fact out of control. This fact can {\bf invalidate}
any attempt for defining the usual (Fock) scattering
states that are asymptotically free for Higgs fields
in any QFT (with Higgs sector) in which the EFP
approach is consistent. That is,

{\bf Proposition II}\ \ \
{\sl The Higgs modes, for which the usual asymptotic
Fock states can not be defined for the full theory,
could not be described by the normal particle concept
and hence it is not possible to identify them
experimentally as normal particles like leptons,
hadrons, intermediate gauge bosons ($W^+, W^-, Z$), etc.}

Thus the Higgs modes are quite different from the
normal particles or field quanta. The Higgs quanta
might be identified otherwise or indirectly. As this
is only an investigation basing on one technical
approach, we hope the observation here might
draw attention to the peculiar aspect of the Higgs
sector or the scalar sector triggering SSB. Given
the recent result of Higgs mass range
($ m_{H}=115^{+116}_{-66} GeV/c^2$, \cite {Fit}), we
should be reminded of the other properties and/or
aspects of the SSB in addition to the conventional
wisdoms. Of course the true detailed properties
of the Higgs modes are beyond our reach presently.
Here we do not attribute the pecularity to the
convexity of EFP as it is a natural property for any
QFT in which EFP can be consistently defined in terms
of stable states or modes. We feel that it is in fact
the SSB that makes the Higgs modes so complicated. Of
course, it can be implied from the above conclusion
that the quantum fluctuations in a QFT with Higgs
sector would be quite different from the ones without
SSB. There must be some unknown microscopic dynamical
mechanisms that lead to SSB and the abnormal "confinement" 
of Higgs "particles" as well as the appearance of the
strange $A^0$ subsector.

It is necessary to emphasize again that since we did
not specify the details of the Higgs sector, our main
results should be applicable to any model containing
a Higgs sector whether the Higgs fields are elementary
or not. We also need to note that though the Higgs
modes might not show up normally they {\sl are}
actively participating all the relevant particle
processes and reactions in the intermediate stages
(the virtual quantum fluctuations). Maybe their
contributions to the physical world were
"appropriated" by the other particles that finally
come into "our macroscopic sights".

Before closing we would like to mention that the triviality
problem \cite {Trivia} associated with the scalar Higgs
model does not affect our arguments here at all. This
is because we need not here assume that Higgs model as
well as SM is the true fundamental theory. In fact, in
Ref. \cite {Yang}, it is clearly demonstrated that given
the standard point of view that a well defined fundamental
theory underlying all the QFTs beset with various
unphysical infinities (esp. UV divergence) there is a
very natural and simple way to calculate radiative
corrections without incurring UV infinities in these
QFTs--a substitute for the conventional
renormalization program with more physical rationality
and almost no mathematical absurdity. The main point
there is to admit first of all that {\sl all the
theories beset with UV divergence are in fact
ill-defined low energy effective theories for the
sectors of the true underlying one.} The so-called
triviality of a QFT is in fact equivalent to saying
that this theory is an {\bf effective} one rather than a 
fundamental one. That theory is only valid below a 
certain energy scale which can not be taken to UV infinity.
However, different from the triviality literature, we hold
that a so-called trivial theory IS nontrivial and hence
useful given that one works with physical parameters
(say, external momenta) below the defining energy scale.
(For detailed discussion, see Ref. \cite {Yang}.) So,
a "trivial" theory is quite useful for physical purposes,
just like any other theory (say, QCD) beset with UV
infinities.

In summary, we suggested a peculiar aspect of the Higgs
sector basing on a rather general property, i.e., the
convexity, of the full effective potential of the scalar
sector together with the requirement of spontaneous
symmetry breaking. It seems to be an alternative way of
thinking about the Higgs physics or even the standard model
physics.
\section*{Acknowledgement}
The author is grateful to Dr. G.-h. Yang, Prof. G-j Ni and
Prof. J-j Xu for helpful discussions.

\end{document}